%% file: MSOS-continuations.tex
\documentclass[copyright,creativecommons]{eptcs}
\providecommand{\event}{WoC 2015} 

\title{A Modular Structural Operational Semantics\\ for Delimited Continuations}
\author{Neil Sculthorpe
\institute{\textsc{PLanCompS} Project\\ Department of Computer Science\\ Swansea University, UK}
\email{N.A.Sculthorpe@swansea.ac.uk}
\and
Paolo Torrini
\institute{\textsc{GRACeFUL} Project\\Department of Computer Science\\ KU Leuven, Belgium}
\email{Paolo.Torrini@cs.kuleuven.be}
\and
Peter D. Mosses
\institute{\textsc{PLanCompS} Project\\ Department of Computer Science\\ Swansea University, UK}
\email{P.D.Mosses@swansea.ac.uk}
}
\def\titlerunning{A Modular Structural Operational Semantics for Delimited Continuations}
\def\authorrunning{Neil Sculthorpe, Paolo Torrini \& Peter D. Mosses}

\usepackage{mathtools} 
\usepackage{amssymb}
\usepackage{continuations}

\usepackage{latexsym}
\usepackage{stmaryrd}

\usepackage{comment}

\usepackage{ifthen}
\newboolean{shortversion}
\setboolean{shortversion}{true}

\newcommand{\lambdaV}{$\lambda\!$V}
\newcommand{\lambdaDC}{$\lambda\!$DC}

\newcommand{\escmathcom}[3]{ \newcommand{#1}[#2]{\mbox{#3}}}
\escmathcom{\mcal}{1}{$\mathcal{#1}$}
\escmathcom{\msf}{1}{$\mathsf{#1}$}

\newcounter{prop}
\renewcommand{\theprop}
{\arabic{prop}}

\newenvironment{propA}
{\refstepcounter{prop}%
 \begin{description}%
 \item[\textbf{Prop.}
       \theprop]}%
{\end{description}}

\newcounter{def}
\renewcommand{\thedef}
{\arabic{def}}

\newenvironment{defA}
{\refstepcounter{def}%
 \begin{description}%
 \item[\textbf{Def.}
       \thedef]}%
{\end{description}}

\begin{document}
\maketitle

\begin{abstract}
\input{abstract.tex}
\end{abstract}

\input{introduction.tex}
\input{delimited.tex}
\input{msos.tex}
\input{dynamic.tex}

\input{adequacy.tex}

\ifthenelse{\boolean{shortversion}}%
{}
{\input{static.tex}}

\ifthenelse{\boolean{shortversion}}%
{
\section{Related Work}
\label{sec:related}
\input{related-dynamic.tex}
}
{}

\input{conclusions.tex}

\input{acknowledgement.tex}

\bibliographystyle{eptcs}
\bibliography{continuations.bib}

\end{document}

%% file: abstract.tex
\ifthenelse{\boolean{shortversion}}%
{It has been an open question as to whether the Modular Structural Operational Semantics framework can express the dynamic semantics of \op{call/cc}.
This paper shows that it can, and furthermore, demonstrates that it can express the more general delimited control operators \op{control} and \op{shift}.
}%
{It has been an open question as to whether the Modular Structural Operational Semantics framework can express the semantics of \op{call/cc}.
This paper show that it can, and furthermore, demonstrates that it can express the more general delimited control operators \op{control} and \op{shift}.
We present both a dynamic and static semantics.
}%

%% file: introduction.tex
\section{Introduction}

Modular Structural Operational Semantics (\MSOS) \cite{Mosses:02:PragmaticMSOS,Mosses:04:MSOS,Mosses:09:IMSOS} is a variant of the well-known Structural Operational Semantics (SOS) framework~\cite{Plotkin:81:SOS}.
The principal innovation of \MSOS{} relative to SOS is that it allows the semantics of a programming construct to be specified independently of any semantic entities with which it does not directly interact.
For example, function application can be specified by \MSOS{} rules without mentioning stores or exception propagation.

While it is known that MSOS can specify the semantics of programming constructs for exception handling~\cite{Churchill:13:Bisimulation,Churchill:15:ReusableComponents,Mosses:02:PragmaticMSOS}, it has been unclear whether MSOS can specify more complex control-flow operators, such as \op{call/cc}~\cite{Abelson:Scheme:98,Clinger:87:SchemeContinuations}.
Indeed, the perceived difficulty of handling control operators has been regarded as one of the main limitations of MSOS relative to other modular semantic frameworks (e.g.\ \cite[Section~2]{Rosu:10:KFramework}).
This paper demonstrates that the \ifthenelse{\boolean{shortversion}}%
{dynamic}
{}
semantics of \op{call/cc} \emph{can} be specified in MSOS,
with no extensions to the MSOS framework required.
We approach this by first specifying the more general delimited control operators
\op{control}~\cite{Felleisen:88:AbstractContinuations,Felleisen:88:FirstClassPrompts,Sitaram:90:ControlDelimitersHierarchies}
and \op{shift}~\cite{Danvy:89:TypedContexts,Danvy:90:AbstractingControl,Danvy:92:RepresentingControl},
and then specifying \op{call/cc} in terms of \op{control}.
In contrast to most other operational specifications of control operators given in direct style (e.g.\ \cite{Felleisen:88:AbstractContinuations,Gunter:95:GeneralisationOfControlInML,Kameyama:08:TypedPromptControl,Shan:07:StaticSimulation}), ours is based on labelled transitions, rather than on evaluation contexts.

\ifthenelse{\boolean{shortversion}}%
{}
{In summary, the contributions of this paper are:

\begin{itemize}

\item
We answer (affirmatively) the question of whether the MSOS framework can express the semantics of control operators.

\item
We present a dynamic small-step semantics for the control operators \op{control}, \op{shift} and \op{call/cc} in the MSOS framework.  (Section~\ref{sec:dynamic})

\item
We present a static semantics for those control operators in the MSOS framework.  (Section~\ref{sec:static})

\end{itemize}
}

We will begin by giving a brief overview of delimited continuations (Section~\ref{sec:delimited}) and MSOS (Section~\ref{sec:msos}).
The material in these two sections is not novel, and can be skipped by a familiar reader.
We will then present our MSOS specification of the dynamic semantics of delimited control operators (Section~\ref{sec:dynamic}).
To ensure that our MSOS specification does indeed define the same control operators as described in the literature, we provide a proof of equivalence between our specification and one based on evaluation contexts (Section~\ref{sec:adequacy}).

%% file: delimited.tex
\section{Delimited Continuations}
\label{sec:delimited}

At any point in the execution of a program, the \emph{current continuation} represents the rest of the computation.
In a meta-language sense, a continuation can be understood as a context in which a program term can be evaluated.
\emph{Control operators} allow the current continuation to be treated as an object in the language, by reifying it as a first-class abstraction that can be applied and manipulated.
The classic example of a control operator is \op{call/cc}~\cite{Abelson:Scheme:98,Clinger:87:SchemeContinuations}.

\emph{Delimited continuations} generalise the notion of a continuation to allow representations of partial contexts, relying on a distinction between inner and outer context.
Control operators that manipulate delimited continuations are always associated with \emph{control delimiters}.
The most well-known delimited control operators are
     \op{control} (associated with the \op{prompt} delimiter)~\cite{Felleisen:88:AbstractContinuations,Felleisen:88:FirstClassPrompts,Sitaram:90:ControlDelimitersHierarchies}
 and \op{shift} (associated with the \op{reset} delimiter)~\cite{Danvy:89:TypedContexts,Danvy:90:AbstractingControl,Danvy:92:RepresentingControl}, both of which can be used to simulate \op{call/cc}.
The general idea of \op{control} and \op{shift} is to capture the current continuation up to the innermost enclosing delimiter, representing the inner context.
We will give an informal description of \op{control} in this section.
The formal \MSOS{} specification of \op{control} is given in Section~\ref{sec:dynamic}, where we also specify \op{shift} and \op{call/cc} in terms of \op{control}.


\op{control} is a (call-by-value) unary operator that takes a higher-order function $f$ as its argument, where $f$ expects a reified continuation as its argument.
When executed, \op{control} reifies the current continuation, up to the innermost enclosing \op{prompt}, as a function $k$.
That inner context is then discarded and replaced with the application $f \; k$.
Other than its interaction with \op{control}, \op{prompt} is simply a unary operator that evaluates its argument and returns the resulting value.

Let us consider some examples.
In the following expression, the continuation $k$ is bound to the function $(\lam{x} 2 * x)$, the result of the \op{prompt} application is $14$, and the expression evaluates to $15$:
\begin{codesize}
\begin{equation*}
1 + \op{prompt} (2 * \op{control} (\lam{k} k \; 7)) \quad \leadsto \quad 15
\end{equation*}
\end{codesize}%
A reified continuation can be applied multiple times, for example:
\begin{codesize}
\begin{equation*}
1 + \op{prompt} (2 * \op{control} (\lam{k} k (k \; 7))) \quad \leadsto \quad 29
\end{equation*}
\end{codesize}%
Furthermore, a continuation need not be applied at all.
For example, in the following expression, the multiplication by two is discarded:
\begin{codesize}
\begin{equation*}
1 + \op{prompt} (2 * \op{control} (\lam{k} 7)) \quad \leadsto \quad 8
\end{equation*}
\end{codesize}%

In the preceding examples, the continuation $k$ could have been computed statically.
However, in general, the current continuation is the context at the point in a program's execution when \op{control} is executed, by which time some of the computation in the source program may already have been performed.
For example, the following program will print \textit{ABB}:
\begin{codesize}
\begin{equation*}
\op{prompt} (\; \mathit{print} \; 'A' \; ; \; \op{control} (\lam{k} (k \; () \; ; \; k \; ())) \; ; \; \mathit{print} \; 'B' \;) \quad \rightsquigarrow \quad \textit{ABB}
\end{equation*}
\end{codesize}%
The command $(\mathit{print} \; 'A')$ is executed before the \op{control} operator, so does not form part of the continuation reified by \op{control}.
In this case, $k$ is bound to $(\lam{x} (x \; ; \; \mathit{print} \; 'B'))$, and so $B$ is printed once for every application of $k$.

Further examples of \op{control} can be found in the online test suite accompanying this paper~\cite{AdditionalMaterial}, and in the literature~\cite{Felleisen:88:AbstractContinuations,Felleisen:88:FirstClassPrompts}.

%% file: msos.tex
\section{Modular SOS}
\label{sec:msos}

The rules in this paper will be presented using \emph{Implicitly Modular SOS} (\IMSOS{})~\cite{Mosses:09:IMSOS}, a variant of \MSOS{} that has a notational style similar to conventional SOS.
\IMSOS{} can be viewed as syntactic sugar for \MSOS{}.
We assume the reader is familiar with SOS (e.g.\ \cite{Astesiano:91:OpSem,Plotkin:81:SOS}) and the basics of \MSOS{}~\cite{Mosses:02:PragmaticMSOS,Mosses:04:MSOS,Mosses:09:IMSOS}.

The key notational convenience of \IMSOS{} is that any semantic entities (e.g.\ stores or environments) that are not mentioned in a rule are \emph{implicitly propagated} between the premise(s) and conclusion, allowing entities that do not interact with the programming construct being specified to be omitted from the rule.
Two types of semantic entities are relevant to this paper:
   inherited entities (e.g.\ environments), which, if unmentioned, are implicitly propagated from the conclusion to the premises,
   and observable entities (emitted signals, e.g.\ exceptions), which, if unmentioned, are implicitly propagated from a sole premise to the conclusion.
Observable entities are required to have a default value, which is implicitly used in the conclusion of rules that lack a premise and do not mention the entity.
Note that by \emph{premise}, we refer specifically to a transition of the $\rightarrow$ relation, not any side conditions on the rule (which, for notational convenience, we also write above the line).

To demonstrate the specification of control operators using \IMSOS{} rules, this paper will use the \emph{funcon framework}~\cite{Churchill:15:ReusableComponents}.
This framework contains an open collection of modular \emph{fundamental constructs} (funcons), each of which has its semantics specified independently by \IMSOS{} rules.
Funcons facilitate formal specification of programming languages by serving as a target language for a specification given by an inductive translation, in the style of denotational semantics.
However, this paper is not concerned with the translation of control operators from any specific language: our aim is to give \MSOS{} specifications of control operators, and the funcon framework is a convenient environment for specifying prototypical control operators.
Examples of translations into funcons can be found in~\cite{Churchill:15:ReusableComponents,Mosses:14:FunKons}.%

We will now present some examples of funcons, and their specifications as small-step \IMSOS{} rules.
No familiarity with the funcon framework is required: for the purposes of understanding this paper the funcons may simply be regarded as abstract syntax.
We typeset funcon names in \fcon{bold}, meta-variables in \textit{Capitalised Italic}, and the names of semantic entities in \ent{sans-serif}.
When we come to funcons for control operators, we will continue to use \emph{italic} when referring to the control operator in general, and \fcon{bold} when referring to the funcon specifically.

\input{throwcatch-rules.tex}

Figure~\ref{fig:exceptions} presents
\IMSOS{} rules for
\ifthenelse{\boolean{shortversion}}%
{}%
{the dynamic semantics of }%
the exception-handling funcons \fcon{throw} and \fcon{catch}~\cite{Churchill:15:ReusableComponents}.
The idea is that \fcon{throw} emits an exception signal, and \fcon{catch} detects and handles that signal.
The first argument of \fcon{catch} is the expression to be evaluated, and the second argument (a function) is the exception handler.
Exception signals use an observable entity named \ent{exc}, which is written as a label on the transition arrow.
The \ent{exc} entity has either the value \fcon{none}, denoting the absence of an exception, or $\fcon{some}(V)$, denoting the occurrence of an exception with value $V$.
The side condition $\predicate{val}{V}$ requires the term $V$ to be a value, thereby controlling the order in which the rules can be applied.
In the case of \fcon{throw}, first the argument is evaluated to a value (Rule~\ref{rule:throw-cong}), and then an exception carrying that value is emitted (Rule~\ref{rule:throw-emit}).
In the case of \fcon{catch}, the first argument $E$ is evaluated while no exception occurs (Rule~\ref{rule:catch-eval}).
If an exception does occur, then the handler $H$ is applied to the exception value and the computation $E$ is abandoned (Rule~\ref{rule:catch-handle}).
If $E$ evaluates to a value $V$, then $H$ is discarded and $V$ is returned (Rule~\ref{rule:catch-val}).

\input{lambda-rules.tex}

Observe that rules \ref{rule:throw-cong} and \ref{rule:catch-val} do not mention the \ent{exc} entity.
In Rule~\ref{rule:throw-cong} it is implicitly propagated from premise to conclusion, and in Rule~\ref{rule:catch-val} it implicitly has the default value \fcon{none}.
Also observe that none of the rules in Figure~\ref{fig:exceptions} mention any other entities such as environments or stores; any such entities are also implicitly propagated.

Figure~\ref{fig:lambda} presents
\IMSOS{} rules for
\ifthenelse{\boolean{shortversion}}%
{}%
{the dynamic semantics of }%
identifier lookup (\fcon{bv}, ``bound-value''),
abstraction (\fcon{lambda}), and
application (\fcon{apply}).
Note that the \fcon{closure} funcon is a \emph{value constructor}~\cite{Churchill:13:Bisimulation}
  (specified by Rule~\ref{rule:closure-val}),
   and thus has no transition rules of its own.
We present these rules here for completeness, as these funcons will be used when defining the semantics of control operators in Section~\ref{sec:dynamic}.

\input{sos-rules.tex}

Again, observe that rules \ref{rule:apply-cong1}--\ref{rule:apply-closure-done} do not mention the environment \ent{env}; it is propagated implicitly.
Furthermore, consider that none of the rules in Figure~\ref{fig:exceptions} mention the environment \ent{env}, and none of the rules in Figure~\ref{fig:lambda} mention the \ent{exc} signal.
However, the modular nature of \IMSOS{} specifications allows the two sets of rules to be combined without modification, with implicit propagation handling the unmentioned entities.
For comparison, in Figure~\ref{fig:sos} we present a conventional SOS specification of this call-by-value lambda calculus combined with exception handling, in which both semantic entities are mentioned explicitly in every rule.

%% file: throwcatch-rules.tex

\begin{figure}[b]
\beforebotfigurespace
\begin{codesize}
\begin{minipage}{0.44\textwidth}
\begin{gather}
\frac
{E \to E'}
{\fcon{throw}(E) \to \fcon{throw}(E')}
\label{rule:throw-cong}
\\[1ex]
\frac
{\predicate{val}{V}}
{\fcon{throw}(V) \xrightarrow{\entity{exc}{\fcon{some}(V)}} \fcon{stuck}}
\label{rule:throw-emit}
\end{gather}
\end{minipage}
\hfill
\begin{minipage}{0.53\textwidth}
\begin{gather}
\frac
{E \xrightarrow{\entity{exc}{\fcon{none}}} E'}
{\fcon{catch}(E,H) \xrightarrow{\entity{exc}{\fcon{none}}} \fcon{catch}(E',H)}
\label{rule:catch-eval}
\\[1ex]
\frac
{E \xrightarrow{\entity{exc}{\fcon{some}(V)}} E'}
{\fcon{catch}(E,H) \xrightarrow{\entity{exc}{\fcon{none}}} \fcon{apply}(H,V)}
\label{rule:catch-handle}
\\[1ex]
\frac
{\predicate{val}{V}}
{\fcon{catch}(V,H) \to V}
\label{rule:catch-val}
\end{gather}
\end{minipage}
\end{codesize}
\ifthenelse{\boolean{shortversion}}%
{\caption{\IMSOS{} rules for exception handling.}}
{\caption{\IMSOS{} rules for the dynamic semantics of exception handling.}}
\label{fig:exceptions}
\afterfigurespace
\end{figure}

%% file: lambda-rules.tex

\begin{figure}[t]
\beforefigurespace
\begin{codesize}
\begin{minipage}{0.49\textwidth}
\begin{gather}
\frac
{\rho(I) = V}
{\entity{env}{\rho} \vdash \fcon{bv}(I) \to V}
\label{rule:bv-lookup}
\\[3ex]
\entity{env}{\rho} \vdash \fcon{lambda}(I,E) \to \fcon{closure}(\rho,I,E)
\label{rule:lambda}
\\[3ex]
\predicate{val}{\fcon{closure}(\rho,I,E)}
\label{rule:closure-val}
\end{gather}
\end{minipage}
\hfill
\begin{minipage}{0.47\textwidth}
\begin{gather}
\frac
{E_1 \to E_1'}
{\fcon{apply}(E_1,E_2) \to \fcon{apply}(E_1',E_2)}
\label{rule:apply-cong1}
\\[1ex]
\frac
{\predicate{val}{V} \also E \to E'}
{\fcon{apply}(V,E) \to \fcon{apply}(V,E')}
\label{rule:apply-cong2}
\\[1ex]
\frac
{\predicate{val}{V_1} \also \predicate{val}{V_2}}
{\fcon{apply}(\fcon{closure}(\rho,I,V_1),V_2) \to V_1}
\label{rule:apply-closure-done}
\end{gather}
\end{minipage}
\begin{gather}
\frac
{\predicate{val}{V} \also \entity{env}{(\{I \mapsto V\} /\rho)}  \vdash E \to E'}
{\entity{env}{\_} \vdash \fcon{apply}(\fcon{closure}(\rho,I,E),V) \to \fcon{apply}(\fcon{closure}(\rho,I,E'),V)}
\label{rule:apply-closure-body}
\end{gather}
\vspace{-3ex}
\ifthenelse{\boolean{shortversion}}%
{\caption{\IMSOS{} rules for call-by-value lambda calculus.}}%
{\caption{\IMSOS{} rules for the dynamic semantics of call-by-value lambda calculus.}}%
\label{fig:lambda}
\end{codesize}
\afterfigurespace
\end{figure}

%% file: sos-rules.tex
\newcommand{\envRho}{\entity{env}{\rho} \vdash}
\newcommand{\excX}{\entity{exc}{X}}
\newcommand{\excNone}{\entity{exc}{\fcon{none}}}

\begin{figure}[b]
\beforebotfigurespace
\begin{codesize}
\begin{minipage}{0.36\textwidth}
\begin{gather*}
\frac
{\envRho E \xrightarrow{\excX} E'}
{\envRho \fcon{throw}(E) \xrightarrow{\excX} \fcon{throw}(E')}
\\[1ex]
\frac
{\predicate{val}{V}}
{\envRho \fcon{throw}(V) \xrightarrow{\entity{exc}{\fcon{some}(V)}} \fcon{stuck}}
\\[4ex]
\frac
{\envRho E \xrightarrow{\entity{exc}{\fcon{none}}} E'}
{\envRho \fcon{catch}(E,H) \xrightarrow{\entity{exc}{\fcon{none}}} \fcon{catch}(E',H)}
\\[1ex]
\frac
{\envRho E \xrightarrow{\entity{exc}{\fcon{some}(V)}} E'}
{\envRho \fcon{catch}(E,H) \xrightarrow{\entity{exc}{\fcon{none}}} \fcon{apply}(H,V)}
\\[1ex]
\frac
{\predicate{val}{V}}
{\envRho \fcon{catch}(V,H) \xrightarrow{\excNone} V}
\end{gather*}
\end{minipage}
\hfill
\begin{minipage}{0.48\textwidth}
\begin{gather*}
\frac
{\rho(I) = V}
{\envRho \fcon{bv}(I) \xrightarrow{\excNone} V}
\\[2ex]
\envRho \fcon{lambda}(I,E) \xrightarrow{\excNone} \fcon{closure}(\rho,I,E)
\\[2ex]
\frac
{\envRho E_1 \xrightarrow{\excX} E_1'}
{\envRho \fcon{apply}(E_1,E_2) \xrightarrow{\excX} \fcon{apply}(E_1',E_2)}
\\[1ex]
\frac
{\predicate{val}{V} \also \envRho E \xrightarrow{\excX} E'}
{\envRho \fcon{apply}(V,E) \xrightarrow{\excX} \fcon{apply}(V,E')}
\\[1ex]
\frac
{\predicate{val}{V} \also \entity{env}{(\{I \mapsto V\} /\rho)}  \vdash E \xrightarrow{\excX} E'}
{\splitfrac{\entity{env}{\_} \vdash \fcon{apply}(\fcon{closure}(\rho,I,E),V) \xrightarrow{\excX}}
           {\fcon{apply}(\fcon{closure}(\rho,I,E'),V)}}
\\[1ex]
\frac
{\predicate{val}{V_1} \also \predicate{val}{V_2}}
{\envRho \fcon{apply}(\fcon{closure}(\rho,I,V_1),V_2) \xrightarrow{\excNone} V_1}
\end{gather*}
\end{minipage}
\end{codesize}
\caption{SOS rules for lambda calculus with exception handling.}
\label{fig:sos}
\afterfigurespace
\end{figure}

%% file: dynamic.tex
\ifthenelse{\boolean{shortversion}}%
{\section{I-MSOS Specifications of Control Operators}}
{\section{Dynamic Semantics}}
\label{sec:dynamic}

We now present a dynamic semantics for control operators in the MSOS framework.
We specify \op{control} and \op{prompt} directly, and then specify \op{shift}, \op{reset} and \op{call/cc} in terms of \op{control} and \op{prompt}.
Our approach is signal-based in a similar manner to the \IMSOS{} specifications of exceptions (Figure~\ref{fig:exceptions}): a control operator emits a signal when executed, and a delimiter catches that signal and handles it.
Note that there is no implicit top-level delimiter around a funcon program---a translation to funcons from a language that does have an implicit top-level delimiter should insert an \emph{explicit} top-level delimiter.

\subsection{Overview of our Approach}
\label{sec:overview}

Whether the semantics of control operators can be specified using MSOS has been considered an open problem (\cite[Section~2]{Rosu:10:KFramework}).
We suspect that this is because there is no explicit representation of a term's context in the MSOS framework---any given rule only has access to the current subterm and the contents of any semantic entities---so it is not immediately obvious how to capture the context as an abstraction.

Our approach is to construct the current continuation of a control operator in the rule for its enclosing delimiter.
We achieve this by exploiting the way that a small-step semantics, for each step of computation, builds a derivation tree from the root of the program term to the current operation.
Thus, for any step at which a control operator is executed, not only will a rule for the control operator be part of the derivation, but so too will a rule for the enclosing delimiter.
At each such step, the current continuation corresponds to an abstraction of the control operator (and its argument) from the subterm of the enclosing delimiter, and thus can be constructed from that subterm.

We represent reified continuations as first-class abstractions, using the \fcon{lambda} funcon from Section~\ref{sec:msos}.
Constructing the abstraction is achieved in two stages: the rule for $\fcon{control}$ replaces the occurrence of $\fcon{control}$ (and its argument) with a fresh identifier, and the rule for $\fcon{prompt}$ constructs the abstraction from the updated subterm.
At a first approximation, this suggests the following rules:
%
\begin{codesize}
\begin{gather}
\frac
{\predicate{fresh-id}{I}}
{\fcon{control}(F) \xrightarrow{\entity{ctrl}{\fcon{some}(F,I)}} \fcon{bv}(I)}
\label{rule:control:approx}
\\[1ex]
\frac
{E  \xrightarrow{\entity{ctrl}{\fcon{some}(F,I)}} E'  \also  K = \fcon{lambda}(I,E')}
{\fcon{prompt}(E) \xrightarrow{\entity{ctrl}{\fcon{none}}} \fcon{prompt}(\fcon{apply}(F,K))}
\label{rule:prompt:approx}
\end{gather}
\end{codesize}%

The side condition $\predicate{fresh-id}{I}$ requires that the identifier $I$ introduced by this rule does not already occur in the program.
Rule \ref{rule:control:approx} replaces the term $\fcon{control}(F)$ with $\fcon{bv}(I)$, and emits a signal ($\ent{ctrl}$) containing the function~$F$ and the identifier $I$.
The signal is then caught and handled by \fcon{prompt} in Rule~\ref{rule:prompt:approx}.
The abstraction $K$ representing the continuation of the executed control operator is constructed by combining $I$ with the updated subterm $E'$ (which will now contain $\fcon{bv}(I)$ in place of $\fcon{control}(F)$).

\subsection{The Auxiliary Environment}

There is one problem with the approach we have just outlined, which is that the identifier $I$ is introduced dynamically when the control operator executes, by which time closures may have already formed.
In particular, if \fcon{control} occurs inside the body of a $\fcon{lambda}$, and the enclosing \fcon{prompt} is outside that $\fcon{lambda}$, then the $\fcon{bv}(I)$ funcon would be introduced inside a $\fcon{closure}$ that has already formed, and hence does not contain a binding for $I$.
For example, consider the evaluation of the following term:
\input{env-problem}The occurrence of $\fcon{bv}(i)$ is now inside a closure containing an empty environment.
Were that closure to be applied (say if this was a subterm of a larger program), then the body of the closure would get stuck, as Rule \ref{rule:apply-closure-body} would provide an environment containing only $x$, which Rule \ref{rule:bv-lookup} could not match.

This problem arises as a consequence of our choice to specify the semantics of lambda calculus using environments and closures.
If we had instead given a semantics using substitution, then this problem would not have arisen.
However, we prefer to use environments because they enable a more modular specification: a substitution-based semantics requires substitution to be defined over every construct in the language.
Moreover, environments allow straightforward semantics for dynamic scope.

Our solution is to introduce an auxiliary environment that is not captured in closures.
Figure~\ref{fig:aux-env} specifies $\fcon{aux-bv}(I)$, which looks up the identifier $I$ in this auxiliary environment, and $\fcon{aux-let-in}(I,V,E)$, which binds the identifier $I$ to the value $V$ in the auxiliary environment and scopes that binding over the expression $E$.
We make use of these funcons in the next subsection, where we give our complete specification of \fcon{control} and \fcon{prompt}.

\input{meta-rules.tex}

\subsection{Dynamic Semantics of \op{control} and \op{prompt}}
\label{sec:control}

We specify \fcon{control} as follows:
%
\vspace{-1ex}
\begin{codesize}
\begin{gather}
\frac
{E \to E'}
{\fcon{control}(E) \to \fcon{control}(E')}
\label{rule:control:cong}
\\[1ex]
\frac
{\predicate{val}{F} \also \predicate{fresh-id}{I}}
{\fcon{control}(F) \xrightarrow{\entity{ctrl}{\fcon{some}(F,I)}} \fcon{aux-bv}(I)}
\label{rule:control:emit}
\end{gather}
\end{codesize}%
Rule~\ref{rule:control:cong}, in combination with the $\predicate{val}{F}$ premise on Rule~\ref{rule:control:emit}, ensures that the argument function is evaluated to a closure before Rule~\ref{rule:control:emit} can be applied.
Notice that Rule~\ref{rule:control:emit} uses $\fcon{aux-bv}$, in contrast to the preliminary Rule~\ref{rule:control:approx} which used $\fcon{bv}$.

We then specify \fcon{prompt} as follows:
%
\vspace{-1ex}
\begin{codesize}
\begin{gather}
\frac
{\predicate{val}{V}}
{\fcon{prompt}(V) \to V}
\label{rule:prompt:val}
\\[1ex]
\frac
{E \xrightarrow{\entity{ctrl}{\fcon{none}}} E'}
{\fcon{prompt}(E) \xrightarrow{\entity{ctrl}{\fcon{none}}} \fcon{prompt}(E')}
\label{rule:prompt:eval}
\\[1ex]
\frac
{E  \xrightarrow{\entity{ctrl}{\fcon{some}(F,I)}} E'  \also  K = \fcon{lambda}(I,\fcon{aux-let-in}(I,\fcon{bv}(I),E'))}
{\fcon{prompt}(E) \xrightarrow{\entity{ctrl}{\fcon{none}}} \fcon{prompt}(\fcon{apply}(F,K))}
\label{rule:prompt:catch}
\end{gather}
\end{codesize}%
Rule~\ref{rule:prompt:val} is the case when the argument is a value; the \fcon{prompt} is then discarded.
Rule~\ref{rule:prompt:eval} evaluates the argument expression while no \ent{ctrl} signal is being emitted by that evaluation.
Rule~\ref{rule:prompt:catch} handles the case when a \ent{ctrl} signal is detected, reifying the current continuation and passing it as an argument to the function $F$.
Notice that, unlike in the preliminary Rule~\ref{rule:prompt:approx}, $I$ is rebound using $\fcon{aux-let-in}$.

Rules~\ref{rule:control:cong}--\ref{rule:prompt:catch} are our complete \IMSOS{} specification of the dynamic semantics of \fcon{control} and \fcon{prompt}, relying only on the existence of the lambda-calculus and auxiliary-environment funcons from figures \ref{fig:lambda} and \ref{fig:aux-env}.
These rules are modular: they are valid independently of whether the control operators coexist with a mutable store, exceptions, input/output signals, or other semantic entities.
Except for the use of an auxiliary environment, our rules correspond closely to those in specifications of \op{control} and \op{prompt} based on evaluation contexts~\cite{Felleisen:88:AbstractContinuations,Kameyama:08:TypedPromptControl}.
However, our rules communicate between \fcon{control} and \fcon{prompt} by emitting signals, and thus do not require evaluation contexts.
In Section~\ref{sec:adequacy}, we present a proof of equivalence between our specification and a conventional one based on evaluation contexts.

\subsection{Dynamic Semantics of \op{shift} and \op{reset}}
\label{sec:shift}

The \op{shift} operator differs from \emph{control} in that every application of a reified continuation is implicitly wrapped in a delimiter, which has the effect of separating the context of that application from its inner context~\cite{Biernacki:06:Folklore}.
This difference between \emph{control} and \emph{shift} is analogous to that between dynamic and static scoping, insofar as with \emph{shift}, the application of a reified continuation cannot access its context, in the same way that a statically scoped function cannot access the environment in which it is applied.

A \fcon{shift} funcon can be specified in terms of \fcon{control} as follows:
%
\begin{codesize}
\begin{gather}
\frac
{E \to E'}
{\fcon{shift}(E) \to \fcon{shift}(E')}
\\[1ex]
\frac
{\predicate{val}{F} \also \predicate{fresh-id}{K} \also \predicate{fresh-id}{X}}
{\fcon{shift}(F) \to \fcon{control}(\fcon{lambda}(K,\fcon{apply}(F,\fcon{lambda}(X,\fcon{reset}(\fcon{apply}(\fcon{bv}(K),\fcon{bv}(X)))))))}
\end{gather}
\end{codesize}%
The key point is the insertion of the \fcon{reset} delimiter; the rest of the lambda-term is merely an $\eta$-expansion that exposes the application of the continuation $K$ so that the delimiter can be inserted (following~\cite{Biernacki:06:Folklore}).
Given this definition of \fcon{shift}, the \fcon{reset} delimiter coincides exactly with \fcon{prompt}:
%
\begin{codesize}
\begin{gather}
\fcon{reset}(E) \to \fcon{prompt}(E)
\end{gather}
\end{codesize}%

Alternatively, the insertion of the extra delimiter could be handled by the semantics of \fcon{reset} rather than that of \fcon{shift}:
%
\beforegatherspace
\vspace{-1ex}
\begin{codesize}
\begin{gather}
\frac
{\predicate{val}{V}}
{\fcon{reset}(V) \to V}
\label{rule:reset:val}
\\[1ex]
\frac
{E \xrightarrow{\entity{ctrl}{\fcon{none}}} E'}
{\fcon{reset}(E) \xrightarrow{\entity{ctrl}{\fcon{none}}} \fcon{reset}(E')}
\label{rule:reset:eval}
\\[1ex]
\frac
{E  \xrightarrow{\entity{ctrl}{\fcon{some}(F,I)}} E'  \also  K = \fcon{lambda}(I,\fcon{reset}(\fcon{aux-let-in}(I,\fcon{bv}(I),E')))}
{\fcon{reset}(E) \xrightarrow{\entity{ctrl}{\fcon{none}}} \fcon{reset}(\fcon{apply}(F,K))}
\label{rule:reset:catch}
\end{gather}
\end{codesize}%
The only difference between rules~\ref{rule:prompt:val}--\ref{rule:prompt:catch} and rules~\ref{rule:reset:val}--\ref{rule:reset:catch} (other than the funcon names) is the definition of $K$ in Rule \ref{rule:reset:catch}, which here has a delimiter wrapped around the body of the continuation.
Given this definition of \fcon{reset}, the \fcon{shift} operator now coincides exactly with \fcon{control}:
%
\begin{codesize}
\begin{gather}
\fcon{shift}(E) \to \fcon{control}(E)
\label{rule:shift-v2}
\end{gather}
\end{codesize}%
This \IMSOS{} specification in Rules \ref{rule:reset:val}--\ref{rule:shift-v2} is similar to the evaluation-context based specification of \op{shift} and \op{reset} in~\cite[Section 2]{Kameyama:08:TypedPromptControl}.

\subsection{Dynamic Semantics of \op{abort} and \op{call/cc}}
\label{sec:callcc}

The \op{call/cc} operator is traditionally \emph{undelimited}: it considers the current continuation to be the entirety of the rest of the program.
In a setting with delimited continuations, this can be simulated by requiring there to be a single delimiter, and for it to appear at the top-level of the program.
Otherwise, the two distinguishing features of \op{call/cc} relative to \op{control} and \op{shift} are first that an applied continuation never returns, and second that if the body of \op{call/cc} does not invoke a continuation, then the current continuation is applied to the result of the \op{call/cc} application when it returns.

To specify \op{call/cc}, we follow Sitaram and Felleisen~\cite[Section 3]{Sitaram:90:ControlDelimitersHierarchies} and first introduce an auxiliary operator \op{abort}, and then specify \op{call/cc} in terms of \op{control}, \op{prompt} and \op{abort}.
The purpose of \op{abort} is to terminate a computation (up to the innermost enclosing \op{prompt}) with a given value:
%
\begin{codesize}
\begin{gather}
\frac
{E \to E'}
{\fcon{abort}(E) \to \fcon{abort}(E')}
\\[1ex]
\frac
{\predicate{val}{V} \also \predicate{fresh-id}{I}}
{\fcon{abort}(V) \to \fcon{control}(\fcon{lambda}(I,V))}
\end{gather}
\end{codesize}%
We achieve the first distinguishing feature of \op{call/cc} by placing an \fcon{abort} around any application of a continuation (preventing it from returning a value), and we achieve the second by applying the continuation to the result of the $F$ application (which resumes the current continuation if $F$ returns a value):
%
\begin{codesize}
\begin{gather}
\frac
{E \to E'}
{\fcon{callcc}(E) \to \fcon{callcc}(E')}
\\[1ex]
\frac
{\predicate{val}{F} \also \predicate{fresh-id}{K} \also \predicate{fresh-id}{X}}
{\splitfrac{\fcon{callcc}(F) \to}
           {\fcon{control}(\fcon{lambda}(K,\fcon{apply}(\fcon{bv}(K),\fcon{apply}(F,\fcon{lambda}(X,\fcon{abort}(\fcon{apply}(\fcon{bv}(K),\fcon{bv}(X))))))))}}
\end{gather}
\end{codesize}%

\ifthenelse{\boolean{shortversion}}%
{}
{
\subsection{Other Approaches to Dynamic Semantics}
\label{sec:related-dynamic}

\input{related-dynamic.tex}
}

\subsection{Other Control Effects}

In Section~\ref{sec:msos} we presented a direct specification of exception handling using a dedicated semantic entity.
If \fcon{throw} and \fcon{catch} (Figure~\ref{fig:exceptions}) were used in a program together with the control operators from this section, this would give rise to two sets of independent control effects, each with independent delimiters.
An alternative would be to specify exception handling indirectly in terms of the control operators (e.g.\ following Sitaram and Felleisen~\cite{Sitaram:90:ControlDelimitersHierarchies}), in which case the delimiters and semantic entity would be shared.
\MSOS{} can specify either approach, as required by the language being specified.

Beyond the control operators discussed in this section, further and more general operators for manipulating delimited continuations exist, such as those of the CPS hierarchy~\cite{Danvy:90:AbstractingControl}.
These are beyond the scope of this paper, and remain an avenue for future work.

%% file: env-problem.tex
\begin{codesize}%
\begin{flalign*}
               & \fcon{prompt}(\fcon{lambda}(x,\fcon{control}(\fcon{lambda}(k,\fcon{bv}(k)))))\\
\rightarrow \; &     \qquad \{ \; \textit{by} \; (\ref{rule:lambda}) \; \} \\
               & \fcon{prompt}(\fcon{closure}(\emptyset,x,\fcon{control}(\fcon{lambda}(k,\fcon{bv}(k)))))\\
\rightarrow \; &     \qquad \{ \; \textit{by} \; (\ref{rule:prompt:approx}) \; \} \\
               & \fcon{prompt}(\fcon{apply}(\fcon{lambda}(k,\fcon{bv}(k)),\fcon{lambda}(i,\fcon{closure}(\emptyset,x,\fcon{bv}(i)))))
\end{flalign*}%
\end{codesize}%

%% file: meta-rules.tex
\begin{figure}[h]
\beforefigurespace
\begin{codesize}
\begin{gather}
\frac
{\mu(I) = V}
{\entity{aux-env}{\mu} \vdash \fcon{aux-bv}(I) \to V}
\\[4ex]
\frac
{E_1 \to E'_1}
{\fcon{aux-let-in}(I,E_1,E_2) \to \fcon{aux-let-in}(I,E'_1,E_2)}
\\[1ex]
\frac
{\predicate{val}{V} \also \entity{aux-env}{(\{I \mapsto V\} /\mu)} \vdash E \to E'}
{\entity{aux-env}{\mu} \vdash \fcon{aux-let-in}(I,V,E) \to \fcon{aux-let-in}(I,V,E')}
\\[1ex]
\frac
{\predicate{val}{V_1} \also \predicate{val}{V_2}}
{\fcon{aux-let-in}(I,V_1,V_2) \to V_2}
\end{gather}
\end{codesize}%
\caption{\IMSOS{} rules for bindings in the auxiliary environment.}
\label{fig:aux-env}
\afterfigurespace
\end{figure}


%% file: related-dynamic.tex
A direct way to specify control operators is by giving an operational semantics based on transition rules and first-class continuations.
We have taken this direct approach, though in contrast to most direct specifications of control operators
(e.g.~\cite{Felleisen:88:AbstractContinuations,Gunter:95:GeneralisationOfControlInML,Kameyama:03:Axiomization,Kameyama:08:TypedPromptControl,Rosu:10:KFramework,Shan:07:StaticSimulation})
our approach is based on emitting signals via labelled transitions rather than on evaluation contexts.
Control operators can also be given a denotational semantics by transformation to continuation-passing style (CPS)~\cite{Danvy:89:TypedContexts,Dyvbig:07:MonadicDelimited,Sabry:93:ReasoningAboutCPS,Shan:07:StaticSimulation}, or a lower-level operational specification by translation to abstract-machine code~\cite{Biernacki:06:StaticDynamicExtent,Felleisen:88:FirstClassPrompts}.
At a higher level, algebraic characterisations of control operators have been given in terms of equational theories~\cite{Felleisen:88:AbstractContinuations,Kameyama:03:Axiomization}.






Denotationally, any function can be rewritten to CPS by taking the continuation (itself represented as a function) as an additional argument, and applying that continuation to the value the function would have returned.
A straightforward extension of this transformation~\cite{Danvy:90:AbstractingControl} suffices to express \op{call/cc}, \op{shift} and \op{reset}; however, more sophisticated CPS transformations are needed to express \op{control} and \op{prompt}~\cite{Shan:07:StaticSimulation}.





Felleisen's~\cite{Felleisen:88:FirstClassPrompts} initial specification of \op{control} and \op{prompt} used a small-step operational semantics without evaluation contexts.
However, this specification otherwise differs quite significantly from ours, being based on exchange rules that push \op{control} outwards through the term until it encounters a \op{prompt}.
As an exchange rule has to be defined for every other construct in the language, this approach is inherently not modular.
Later specifications of \op{control} and \op{prompt} used evaluation contexts and algebraic characterisations based on the notion of \emph{abstract continuations}~\cite{Felleisen:88:AbstractContinuations}, where continuations are represented as evaluation contexts and exchange rules are not needed.
Felleisen~\cite{Felleisen:88:FirstClassPrompts} also gave a lower-level operational specification based on the CEK abstract machine, where continuations are treated as frame stacks.

The \op{shift} and \op{reset} operators were originally specified denotationally, in terms of CPS semantics~\cite{Danvy:89:TypedContexts,Danvy:90:AbstractingControl}.
Continuations were treated as functions, relying on the meta-continuation approach~\cite{Danvy:89:TypedContexts} which distinguishes between outer and inner continuations.
Correspondingly, the meta-continuation transformation produces abstractions that take two continuation parameters, which can be further translated to standard CPS.
A big-step style operational semantics for \emph{shift} has been given by Danvy and Yang~\cite{Danvy:99:CPSHeirarchy}, and a specification based on evaluation contexts has been given by Kameyama and Hasegawa~\cite{Kameyama:03:Axiomization}, together with an algebraic characterisation.






Giving a CPS semantics to \op{control} is significantly more complex than for \emph{shift}~\cite{Shan:07:StaticSimulation}.
This is because the continuations reified by \op{shift} are always delimited when applied, and so can be treated as functions, which is not the case for \op{control}.
Different approaches to this problem have been developed, including
   abstract continuations~\cite{Felleisen:88:AbstractContinuations},
   the monadic framework in~\cite{Dyvbig:07:MonadicDelimited},
   and the operational framework in~\cite{Biernacki:06:StaticDynamicExtent}.
Relying on the introduction of recursive continuations, Shan \cite{Shan:07:StaticSimulation} provides an alternative approach based on a refined CPS transform.
Conversely, the difference between \op{control} and \op{shift} can manifest itself quite intuitively in the direct specification of these operators---whether in our \IMSOS{} specifications (Section~\ref{sec:shift}), or in specifications using evaluation contexts~\cite{Felleisen:88:AbstractContinuations,Kameyama:03:Axiomization,Kameyama:08:TypedPromptControl,Shan:07:StaticSimulation}.

As shown by Filinski~\cite{Filinski:94:RepresentingMonads}, \emph{shift} can be implemented in terms of \op{call/cc} and mutable state, and from the point of view of expressiveness, any monad that is functionally
expressible can be represented in lambda calculus with \op{shift} and \op{reset}.
Moreover, \op{control} and \op{shift} are equally expressive in the untyped lambda calculus~\cite{Shan:07:StaticSimulation}.
A direct implementation of \op{control} and \op{shift} has been given by Gasbichler and Sperber~\cite{Gasbichler:02:FinalShift}.
A CPS-based implementation of control operators in a monadic framework has been given by Dyvbig et al~\cite{Dyvbig:07:MonadicDelimited}.
A semantics of \op{call/cc} based on an efficient implementation of evaluation contexts is provided in the K Framework~\cite{Rosu:10:KFramework}.



%% file: adequacy.tex
\section{Adequacy}
\label{sec:adequacy}

Our SOS model of call-by-value lambda calculus extended with delimited
control, which we have presented using I-MSOS rules, is provably
equivalent to one based on the reduction semantics (RS) of lambda
terms where the evaluation strategy is specified using evaluation
contexts.  Reduction models for delimited control based on evaluation
contexts were originally introduced in
\cite{Felleisen:88:AbstractContinuations} and refined in
\cite{Kameyama:08:TypedPromptControl}. The adequacy proof in this
section (Prop.~\ref{prop:final}) is carried out with respect to our
version of those models in a formalism that we call RC.

Our SOS model differs from reduction models in the framework it relies on.
In particular, our SOS model uses environments and signals, whereas RC uses
substitution and evaluation contexts. Moreover, there is a difference
in the notion of value: in our SOS model function application is
computed using closures, whereas RS uses $\beta$-reduction
and substitution. In order to focus on the operational content of the
models, it is convenient to get above these differences. We achieve
this by embedding SOS in RS with explicit congruence rules (an
embedding that we call LS), and by lifting RC to an environment-based
formalism (called LR). We define a notion of adequacy between two
systems, as an input-output relation, parametric in a translation.  We
show adequacy of two systems by proving that they are derivationally
equivalent (in the sense of a step-wise relation), reasoning by
induction on the structure of derivations.  Our adequacy proof for SOS
and RC is split into three main parts: the equivalence of SOS and LS,
of LS and LR, and of LR and RC. A more challenging approach would
involve giving a formal derivation of an RC model from SOS along the
lines of \cite{Danvy:08:Defunctionalized}, but that goes beyond the
scope of this paper.

Here we intend to focus on equivalence with respect to delimited
control. Given the equivalence between SOS and RC with respect to
call-by-value lambda calculus (\lambdaV{}), we show that SOS and RC
are equivalent with respect to the extension of \lambdaV{} with
delimited control (\lambdaDC{}). More specifically, we define a
syntactic representation of environments (standard and auxiliary ones)
using contexts and lambda terms. We use this representation to define
LS as a lambda-term encoding of SOS. Adequacy between SOS and LS is
provable with respect to a simple translation relation.

We define LR as an environment-based version of RC obtained by
lambda-lifting. We consider two distinct extensions of the LR model
of \lambdaV{} with delimited control. The first one, which we call
LR-DC, uses the lifted control rules of the original RC model, and thus
equivalence with the RC model is straightforward. The second one,
which we call LX-DC, uses the LS version of the SOS control rules. The
difference between the LS model of \lambdaDC{} and LX-DC, which are
provably equivalent, boils down to that between SOS transitions, based
on congruence rules and also using closures, and RC transitions, based
on evaluation contexts and using only lambda expressions. The adequacy
of LX-DC and LR-DC, proved with respect to the identity translation
(Prop.~\ref{prop:R-S}), gives us the result of primary interest.

\subsection{Reduction Semantics}

Our presentation of reduction semantics with evaluation contexts (RC)
follows the main lines of the \lambdaDC{} model in
\cite{Kameyama:08:TypedPromptControl}. Under the assumption that we
only evaluate closed expressions, values and terms can be defined as
follows:
\begin{gather}
V \; = \; \msf{lambda}(I,E)  \\[0.4ex]
E \; = \; V \mid \msf{bv}(I) \mid
\msf{apply}(E, E) \mid \msf{control}(E) \mid \msf{prompt}(E)
\end{gather}
A general notion of a context as a term with a hole can be defined by
the following grammar:
\begin{gather}
C \; = \; [] \mid \msf{lambda}(I,C) \mid \msf{apply}(C, E) \mid
\msf{apply}(E, C) \mid \msf{prompt}(C) \mid \msf{control}(C)
\end{gather}
The call-by-value (CBV) evaluation strategy can be specified using the more
restrictive notion of a CBV context ($Q$):
\begin{gather}
Q \; = \; [] \mid \msf{apply}(Q, E) \mid \msf{apply}(V, Q) \mid
\msf{prompt}(Q) \mid \msf{control}(Q)
\end{gather}
In order to represent delimited continuations, an even more
restrictive notion is needed: a \emph{pure} context ($P$-context),
which is a CBV-context that does not include control delimiters
\cite{Kameyama:08:TypedPromptControl}:
\begin{gather}
P \; = \; [] \mid \msf{apply}(P, E) \mid \msf{apply}(V, P) \mid
\msf{control}(P)
\end{gather}
The meta-linguistic notation $C[E]$ ($Q[E]$, $P[E]$) is used to
represent a term factored into a context and the subterm that fills
the hole---we can think of this as a form of term annotation.  This
factorisation is unique for the cases that we are considering.

The only reduction rules needed to specify \lambdaV{} are
$\beta$-reduction and context propagation.  These can be presented as
follows (giving us the RC-V model), using $\{ \_ \mapsto \_ \}$ as
meta-level notation for capture-avoiding uniform substitution:
\begin{gather}
 \msf{apply}(\msf{lambda}(I,E), V) \longrightarrow E \{ \msf{bv}(I)
 \mapsto V \}
\\[1ex]
\frac{ E \longrightarrow E' }
{ Q[E] \longrightarrow Q[E'] }
\end{gather}
The reduction rules for $\msf{prompt}$ and $\msf{control}$ can be
formulated as follows (giving us the RC-DC model), making use of
$P$-contexts:
\begin{gather} \label{EC-prompt}
\msf{prompt}(V) \longrightarrow V
\\[1ex]
\label{EC-control}
\frac{ val(F) \qquad K = \msf{lambda}(I,P[\msf{bv}(I)]) \qquad
  \predicate{fresh-id}{I} } {\msf{prompt}(P[\msf{control}(F)])
  \longrightarrow \msf{prompt}(\msf{apply}(F,K))}
\end{gather}

In a system based on reduction semantics, \emph{observational
  equivalence} can be defined as the smallest congruence relation
$\equiv$ on terms that extends reduction equivalence with functional
extensionality, i.e.~such that
$$
\frac{
\forall \; V. \; \msf{apply}(F,V) \equiv
\msf{apply}(F',V) }
{ F \equiv F' }
$$

In presenting models based on RS, we typeset all construct names in
\textsf{sans-serif}. We refer to SOS values as $Val_{\tiny \mbox{SOS}}$ and to
RC ones as $Val_{\tiny \msf{RC}}$. When needed, we subscript
$\longrightarrow$ and $\equiv$ accordingly.  We define
\emph{derivational equivalence} and \emph{adequacy} with respect to a
translation relation (not mentioned in the case that it is an identity), as
follows.
\begin{defA}
Given two systems $A$ and $B$, respectively defined on languages $L_A$ and
$L_B$ with values $Val_A$ and $Val_B$, and a one-to-one relation $R \subseteq (L_A,L_B)$, we say that
\begin{enumerate}
\item $A$ and $B$ are \emph{adequate} with respect to $R$ ($A \sim^{R} B$) whenever the following hold:\\
A) If $E \longrightarrow^*_A V$, with $V \in Val_A$, then there exist
$E'_a,E''_a \in L_A$, $E'_b, E''_b \in L_B$, $V' \in Val_B$ s.t. $E
\equiv_A E'_a$, $V \equiv_A E''_a$, $R(E'_a,E'_b)$, $R(E''_a,E''_b)$,
$E''_b \equiv_B V'$ and $E'_b \longrightarrow^*_B V'$ \\
B) If $E \longrightarrow^*_B V$, with $V \in Val_B$, then there exist
$E'_a,E''_a \in L_A$, $E'_b, E''_b \in L_B$, $V' \in Val_A$ s.t. $E
\equiv_B E'_b$, $V \equiv_B E''_b$, $R(E'_a,E'_b)$, $R(E''_a,E''_b)$,
$E''_a \equiv_A V'$ and $E'_a \longrightarrow^*_B V'$
%
%
%
\item
$A$ and $B$ are \emph{derivationally equivalent} with respect to
$R$ whenever the following hold:\\
A) $E_1 \longrightarrow_A E_2$ whenever there exist $E_3, \; E_4,
\; E'_1, \; E'_2, \; E'_3, \; E'_4$ s.t.~$E_1 \equiv_A E_3$, $E_2
\equiv_A E_4$, $R(E_3,E'_3)$, $R(E_4, E'_4)$, $E'_3 \equiv_B E'_1$, $
E'_4 \equiv_B E'_2$ and $E'_1\longrightarrow^*_B E'_2$ \\
B) $E_1 \longrightarrow_B E_2$ whenever there exist $E_3, \; E_4,
\; E'_1, \; E'_2, \; E'_3, \; E'_4$ s.t.~$E_1 \equiv_B E_3$, $E_2
\equiv_B E_4$, $R(E'_3,E_3)$, $R(E'_4, E_4)$, $E'_3 \equiv_A E'_1$, $
E'_4 \equiv_A E'_2$ and $E'_1\longrightarrow^*_A E'_2$
%
\end{enumerate}
\end{defA}

We define relational composition as $R_1 \circ R_2 = \lambda x
y. \; \exists z. \; R_1(x,z) \land R_2(z,y)$.

\subsection{Representing SOS as LS}

In this section we define LS, as an encoding of SOS in
lambda terms. Unlike reduction semantics, our SOS models rely
internally on a linguistic extension to account for closures and the
auxiliary environment notation. For this reason, we need an extended
\emph{internal} language, including the following additional
constructs: $\msf{closure}(\rho,I,E)$ for closures,
$\textsf{aux-bv}(I)$ for auxiliary identifier lookup, and
$\textsf{aux-let-in}(I,E,E)$ for auxiliary let bindings; these
constructs are not meant to be included in the source language
definition. In each expression $\textsf{aux-let-in}(I, E, E')$, we
require $I$ to be used at most once in $E'$.

In order to represent environments, we introduce a notion
of $R$-context:
\begin{gather}
R \; = \; [] \mid \msf{apply}(\msf{lambda}(I, R),V)
\end{gather}
We tacitly assume that all bound variables in $R$ are distinct. Each
SOS transition specified by
$$\textsf{env}\;\rho \vdash E \longrightarrow E'$$


\vspace{-2ex}

\noindent
can be embedded as

\vspace{-3ex}

$$ R_{\rho}[E] \longrightarrow R_{\rho}[E'] $$
where, for $\rho = \{ I_1 \mapsto V_1, \ldots, I_n \mapsto V_n \} $,
$R_{\rho} = \msf{apply}(\msf{lambda}
(I_n,(\ldots,\msf{apply}(\msf{lambda}(I_1, []), V_1),\ldots)),V_n )$.
We silently assume permutation in $R$-contexts.
We introduce $M$-contexts to represent the auxiliary environment, in a
similar manner to $R$-contexts, though using \textsf{aux-let-in}. In
order to represent signals, we extend this notion to one of
$S$-context, introducing a new ternary value constructor
\textsf{ctrl}, which is not part of the expression definition but only
of the RS representation of SOS.
\begin{gather}
M \; = \; [] \mid \textsf{aux-let-in}(I, V, M)
\\[0.4ex]
S \; = \; M \mid \textsf{ctrl}(V, I, M)
\end{gather}
%
SOS transitions specified by
$$
\begin{array}{llll}
\mbox{A})  & \textsf{aux-env}\;\mu, \textsf{env}\;\rho \vdash E
\xrightarrow{\entity{ctrl}{\fcon{none}}} E' &
\qquad \qquad  \mbox{B})  & \textsf{aux-env}\;\mu, \textsf{env}\;\rho \vdash E
\xrightarrow{\entity{ctrl}{\fcon{some}(F,I)}} E'
\end{array}
$$
can be represented, respectively, as
$$
\begin{array}{llll}
\mbox{A})  & M_{\mu}[R_{\rho}[E]] \longrightarrow M_{\mu}[R_{\rho}[E']] &
\qquad \qquad  \mbox{B})  & M_{\mu}[R_{\rho}[E]] \longrightarrow
\textsf{ctrl}(F,I,M_{\mu}[R_{\rho}[E']])
\end{array}
$$
where, for $\mu = \{ I_1 \mapsto V_1, \ldots, I_n \mapsto V_n \} $,
$M_{\mu} = \textsf{aux-let-in} (I_n,V_n, \ldots
\textsf{aux-let-in}(I_1, V_1, []))$.
We assume that $S$-bound variables are distinct from each other and
from all the $R$-bound ones. As with $R$-contexts, we silently assume
permutation for $M$-contexts.

In this way, we define a one-to-one translation relation
$\textsf{T}(\_,\_)$ between SOS configurations and LS
expressions. Applying the translation to the SOS rules gives us the LS
models for \lambdaV{} and \lambdaDC{} (resp. LS-V and LS-DC). In
particular, the rule for \textsf{bv}, which uses the environment, can
be expressed as follows:
\begin{gather} \label{rule:bv}
{ S[\msf{apply}(\msf{lambda}(I,R[\msf{bv}(I)]),V)] \longrightarrow
  S[\msf{apply}(\msf{lambda}(I,R[V]),V)] }
\end{gather}
%
%
The \textsf{aux-bv} rule, which uses the auxiliary environment, takes the
following form:
\begin{gather}
 \textsf{aux-let-in}(I, V, M[R[\textsf{aux-bv}(I)]]) \longrightarrow
 \textsf{aux-let-in}(I, V, M[R[V]])
\end{gather}
%


\noindent The LS model for \lambdaDC{} extends LS-V with three more
rules: the SOS rule for \textsf{control},
\begin{gather} \label{controlRSOS}
\frac{ val(F) \also \predicate{fresh-id}{I}} {
  M[R[\msf{control}(F)]] \longrightarrow
  \textsf{ctrl}(F,I,M[R[\textsf{aux-bv}(I)]]) }
\end{gather}
an additional congruence rule (which can only match Rule
\ref{controlRSOS}),
\begin{gather} \label{liftControlRSOS}
\frac{ M[R[E]] \longrightarrow \textsf{ctrl}(F,I,M[R[E']]) }
{ M[R[P[E]]] \longrightarrow
  \textsf{ctrl}(F,I,M[R[P[E']]]) }
\end{gather}
%
and an encoding of the SOS rule for \textsf{control}-in-\textsf{prompt}
(Rule~\ref{rule:prompt:catch}).
\begin{gather} \label{promptRSOS}
\frac{ M[R[E]] \longrightarrow \textsf{ctrl}(F, I', M[R[E']])
  \qquad K = \msf{lambda}(I,\textsf{aux-let-in}(I',\msf{bv}(I), E'))
  \qquad \predicate{fresh-id}{I}
  } { M[R[\msf{prompt}(E)]]
  \longrightarrow M[R[\msf{prompt}(\msf{apply}(F,K)]] }
\end{gather}

Since there is a one-to-one correspondence between LS and SOS transitions
(treating $R$- and $M$-permutations as silent transitions),
and taking for simplicity the identity relation modulo reordering of
the environments as observational equivalence in SOS, the following is
straightforward.


\begin{propA} \label{obseq0}
The LS model of \lambdaV{} and the corresponding SOS one are
derivationally equivalent and adequate with respect to the translation
$T$, and similarly for the LS and SOS models of \lambdaDC{}.
\end{propA}

\noindent Proof: First we prove derivational equivalence, which is
straightforward for \lambdaV{}. The LS \textsf{control} rules
correspond to the SOS ones, given the representation of the auxiliary
environments and signals.
Adequacy follows as the definition of value is the same in all these
systems.


\subsection{Lifting RC to LR}

To facilitate comparison with LS,
we define
LR as an environment-based version of RC, using $R$- and $M$-contexts
to represent environments as in LS, and also extend the language
with \textsf{aux-let-in} and \textsf{closure}. In the LR models, the
reduction rules can be specified as in RC, relying on evaluation
contexts. For the way \textsf{aux-let-in} and \textsf{closure} are
used, no change is needed in the definition of context. However, since
reduction steps now have to be lifted by $R$- and $M$-contexts, we replace the
single context propagation rule that sufficed in RC with the four following rules: \emph{lifting}, \emph{lifted
  congruence} and $R$- and $M$-\emph{lowering}.

\noindent 
\begin{gather}
 \frac{ E \longrightarrow E' }
{ M[R[E]] \longrightarrow M[R[E']] }
\\[1ex]
\frac{ M[R[E]] \longrightarrow M[R[E']] }
{ M[R[Q[E]]] \longrightarrow M[R[Q[E']]] }
\\[1ex]
\label{r-unlift1}
 \frac{ M[\msf{apply}(\msf{lambda}(I,R[E]),V)] \longrightarrow
   M[\msf{apply}(\msf{lambda}(I,R[E']),V)] } {
   M[R[\msf{apply}(\msf{lambda}(I,E),V)]] \longrightarrow
   M[R[\msf{apply}(\msf{lambda}(I,E'),V)]] }
\\[1ex]
\label{m-unlift1}
 \frac{ \textsf{aux-let-in}(I, V, M[R[E]]) \longrightarrow
    \textsf{aux-let-in}(I, V, M[R[E']]) }
   { M[R[\textsf{aux-let-in}(I, V, E)]] \longrightarrow
         M[R[\textsf{aux-let-in}(I, V, E')]]}
\end{gather}
We assume that LR models include $R$- and $M$-permutations, as well
as the \textsf{bv} and \textsf{aux-bv} rules, and Rule~\ref{liftControlRSOS}.
Notice that evaluation can also apply to open
terms, however we do not need to change our definition of value, as
substitution of free variables is dealt with as in LS, by the
\textsf{bv} rule. We also need the following rule to deal with
closures:
\begin{gather}
\label{eta-eq3}
 \frac{ \rho = \{ I_1 \mapsto V_1, \ldots I_n
  \mapsto V_n \} \qquad \mbox{\it free-vars}(E) \subseteq \{I_1,
  \ldots, I_n\} }{ \msf{closure}(\rho,I,E) \longrightarrow
\msf{apply}(\msf{lambda}
  (I_n,(\ldots,\msf{apply}(\msf{lambda}(I_1, \msf{lambda}(I,E)),
  V_1),\ldots)),V_n ) }
\end{gather}

This gives us the LR model of \lambdaV{} (LR-V). We can extend
this model with rules for delimited control in two ways: to
simulate RC (LR-DC), or to simulate SOS (LX-DC).
The LR-DC rules for \textsf{prompt} and \textsf{control}
are those based on the RC one (i.e.~they are the lifted version of
Rules \ref{EC-prompt} and \ref{EC-control}), and they do not involve
any use of the auxiliary notation. LR-DC is the extension of LR-V with
these rules.  On the other hand, the LX-DC model is obtained by
extending LR-V with the LS control rules (Rules \ref{controlRSOS}
and \ref{promptRSOS}, which rely on the auxiliary notation).


The following can be proved for all the systems we are considering,
i.e.~with respect to $\equiv_X$ where $X \in \{ \mbox{LS-V},
\mbox{LR-V}, \mbox{LS-DC}, \mbox{LX-DC}, \mbox{LR-DC} \}$.

\begin{propA} \label{closure-app2}
A) $V \in Val_{\tiny \mbox{SOS}}$ whenever there exists
$V' \in Val_{\tiny \mbox{RC}}$ such that $V \equiv_X V'$. \\
\quad $\mbox{}$ \quad B) $S[R_{\rho}[\msf{apply}(\msf{lambda}(I,E),V_1)]]
\longrightarrow_{\tiny \mbox{LS}} V_2$ with $V_1,V_2 \in Val_{\tiny
  \mbox{SOS}}$ whenever there exist $V_3, V_4 \in Val_{\tiny
  \mbox{RC}}$ such that $S[[\msf{apply}(\msf{closure}(\rho,I,E),V_3)]]
\; \longrightarrow_{\tiny \mbox{LR}} V_4$, with $V_1 \equiv_X V_3$ and
$V_2 \equiv_X V_4$.
\end{propA}

The following provable equivalences correspond respectively to the
\textsf{bv} rule, to $\beta$-reduction, and to \textsf{aux-let-in}
elimination, for $\equiv_X$ as before:
\begin{gather}
\label{extra-mach}
\msf{apply}(\msf{lambda}(I,\msf{bv}(I)),V) \equiv_X
  \msf{apply}(\msf{lambda}(I,V),V)
\\[0.4ex]
\label{rule:beta-conv}
\msf{apply}(\msf{lambda}(I,E), V) \; \equiv_X \; E \{ \msf{bv}(I)
\mapsto V \}
\\[0.4ex]
\label{rule:aux-let-in}
\textsf{aux-let-in}(I, V, E) \equiv_X E \{ \textsf{aux-bv}(I)
\mapsto V \}
\end{gather}

The following can now be proved:

\begin{propA} \label{auxPromptAd00}
$ \msf{apply}(\msf{lambda}(I,\textsf{aux-let-in}(I',
  \msf{bv}(I),P[\textsf{aux-bv}(I')])),V) \; \equiv_X
  \; \msf{apply}(\msf{lambda}(I, P[\msf{bv}(I)]),V) $
\end{propA}

\noindent
Proof:
$\msf{apply}(\msf{lambda}(I,\textsf{aux-let-in}(I',
\msf{bv}(I),P[\textsf{aux-bv}(I')])),V) \; \equiv_X
\; \textsf{aux-let-in}(I',V,P[\textsf{aux-bv}(I')])$,
by Equiv.~\ref{rule:beta-conv}.

$\textsf{aux-let-in}(I',V,P[\textsf{aux-bv}(I')]) \; \equiv_X \; P[V]$,
by Equiv.~\ref{rule:aux-let-in}, observing that $\textsf{aux-bv}(I')$
cannot occur free in $P$, as it must be used at most once in
$P[\textsf{aux-bv}(I')]$.

$P[V] \; \equiv_X \; \msf{apply}(\msf{lambda}(I, P[\msf{bv}(I)]),V)$, by Equiv.~\ref{rule:beta-conv}.

\medskip

The following is an immediate consequence of
Prop.~\ref{auxPromptAd00}, applying functional extensionality.
\begin{gather}
\label{auxPromptAd}
 \msf{lambda}(I,\textsf{aux-let-in}(I',
  \msf{bv}(I),P[\textsf{aux-bv}(I')])) \; \equiv_X
  \; \msf{lambda}(I, P[\msf{bv}(I)])
\end{gather}

\subsection{Adequacy of SOS and RC}

We first show that the LR models and the RC ones are equivalent.

\begin{propA} \label{prop:LR-LS}
LR-V and RC-V are adequate, and so too are LR-DC and RC-DC.
\end{propA}

\noindent
Proof: The language of RC is included in that of LR, hence we can
take the identity on RC (denoted by $\textsf{Id}_{\tiny \mbox{RC}}$)
as the translation. The LR models can be obtained by a lambda-lifting
refactoring of the RC models, and this gives equivalent systems, under a
change of the evaluation strategy that affects only the top level. We
prove derivational equivalence by induction on the structure of
derivations, relying on Equiv.~\ref{rule:aux-let-in} and Rule
\ref{eta-eq3} to eliminate the additional LR syntax, observing that
\textsf{aux-let-in} and \textsf{closure} are inessential in LR-DC (they
can only be eliminated without leading to any new values).  Adequacy
follows immediately as values are defined in the same way in the two
systems.

\medskip

%



%

We consider the relationship between the different representations of
\lambdaV{}.

\begin{propA} \label{prop:lambdaDS}
The LS model of \lambdaV{} and the corresponding LR one are adequate.
%
\end{propA}
%
Proof: The two models use the same language, hence we can take the
identity translation. They differ on congruence rules and reduction of
function application. Congruence rules in LR are expressed using
CBV-contexts, unlike in LS, but both are equivalent specifications of
CBV. For equivalence with respect to values and function application,
we rely on Prop.~\ref{closure-app2}.

\medskip


We extend this result to SOS-style delimited control.

\begin{propA} \label{prop:S-LS}
LS-DC and LX-DC are adequate.
\end{propA}
Proof: We first prove derivational equivalence with respect to
identity using Prop.~\ref{prop:lambdaDS} and the fact that the two
extensions are obtained by adding the same rules.

\medskip








We finally compare SOS-style and RC-style delimited control.

\begin{propA} \label{prop:R-S}
LX-DC and LR-DC are derivationally equivalent and adequate.
\end{propA}

\noindent
Proof: We prove derivational equivalence by induction on the structure
of derivations, with respect to the identity translation. The two
systems are equivalent up to \lambdaV{} by Prop.~\ref{prop:LR-LS},
so the only interesting case is delimited control, in which respect
LX-DC stepwise behaves as LS-DC. The lifted version of the RC
\textsf{prompt} rule (Rule \ref{EC-prompt}) is in both systems.
Rules \ref{controlRSOS} and
\ref{liftControlRSOS} can be added to LR-DC
without expanding the set of derivable values. Thus the
only possible difference between the two systems is
between the natural LR \textsf{control} rule (the lifted
version of Rule~\ref{EC-control}) of LR-DC, and the LS
\textsf{control}-in-\textsf{prompt} rule (Rule \ref{promptRSOS}) of
LX-DC.  We now show that the two rules are interderivable (i.e. given
the system with one rule, the other one is derivable). First we
observe that, by Eq.~\ref{auxPromptAd}, the specification of the
continuation $K$ in either rule is equivalent to that in the other,
and therefore interchangeable.

From S to R: in order to derive the
LR-DC rule from the LX-DC one, we observe that a lifted expression
$M[R[P[\msf{control}(F)]]]$, where $F$ is a value, can be reduced to
$\textsf{ctrl}(F,I,M[R[P[\textsf{aux-bv}(I)]]])$ in LX-DC, using the
\textsf{control} rule (Rule \ref{controlRSOS}), and the applicable
congruence rule (Rule \ref{liftControlRSOS}).  This gives us the
premise for the application of the LX-DC
\textsf{control}-in-\textsf{prompt} rule to
$M[R[\msf{prompt}(P[\msf{control}(F)])]]$ in a way that simulates the
LR-DC \textsf{control} rule.

From R to S: in LX-DC (as in
LS-DC) a one-step transition from $M[R[E]]$ to
$\textsf{ctrl}(F,I,M[R[E']])$ is only possible provided $E \equiv
P[\msf{control}(F)]$ and $E' \equiv P[\textsf{aux-bv}(I)]$ for some
$P$ (possibly relying on the conversion of closures to function
applications). Therefore, the LR-DC \textsf{control} rule can be
applied to $M[R[\msf{prompt}(E)]]$ to simulate Rule
\ref{promptRSOS}.

\medskip

Diagramatically, the overall proof can be presented as follows (where
vertical arrows denote model inclusion).
$$
\begin{array}{lllllllll}
\mbox{SOS-DC}  & \sim^{\textsf{T}} & \mbox{LS-DC}   & \sim^{\textsf{Id}} & \mbox{LX-DC}  & \sim^{\textsf{Id}} & \mbox{LR-DC}   & \sim^{\textsf{Id}_{\tiny \mbox{RC}}} & \mbox{RC-DC} \\
\quad \uparrow &                & \quad \uparrow &                 & \quad \uparrow &                 & \quad \uparrow &                             & \quad \uparrow \\
\mbox{SOS-V}   & \sim^{\textsf{T}} & \mbox{LS-V}    & \sim^{\textsf{Id}} & \mbox{LR-V}   & =               & \mbox{LR-V}     & \sim^{\textsf{Id}_{\tiny \mbox{RC}}} & \mbox{RC-V} \\
\end{array}
$$

\begin{propA} \label{prop:final}
SOS-DC and RC-DC are derivationally equivalent and adequate with
respect to $\textsf{T} \circ \textsf{Id}_{\tiny \mbox{RC}}$
\end{propA}
\noindent Proof: based on Props.~\ref{obseq0}, \ref{prop:LR-LS},
\ref{prop:S-LS}, \ref{prop:R-S}, using the fact that adequacy is
transitive, by composition of the translation relations, and by
transitivity of observational equivalence.





%% file: conclusions.tex
\section{Conclusion}

We have presented a
\ifthenelse{\boolean{shortversion}}%
{dynamic}
{dynamic and static}
 semantics for control operators in the \MSOS{} framework, settling the question of whether \MSOS{} is expressive enough for control operators.
Our definitions are concise and modular, and do not require the use of evaluation contexts.
Definitions based on evaluation contexts are often even more concise than the corresponding MSOS definitions, since a single alternative in a context-free grammar for evaluation contexts subsumes an entire MSOS rule allowing evaluation of a particular subexpression.
However, such grammars are significantly less modular than MSOS rules:
   adding a new control operator to a specified language may require duplication of a (potentially large) grammar \cite[e.g.\ pages 141--142]{Felleisen:09:PLT-Redex}.
(This inherent lack of modularity of evaluation context grammars is addressed in the PLT Redex tools by the use of ellipsis.)

We initially validated our specifications through a suite of 70 test programs, which we accumulated from examples in the literature on control operators (\cite{Abelson:Scheme:98,Asai:07:PolyDelimitedCont,Biernacka:05:DelimitedCPS,Biernacki:06:StaticDynamicExtent,Clinger:87:SchemeContinuations,Danvy:06:AnalyticalApproach,Danvy:89:TypedContexts,Felleisen:88:AbstractContinuations,Felleisen:88:FirstClassPrompts,Gunter:95:GeneralisationOfControlInML,Shan:07:StaticSimulation}).
\ifthenelse{\boolean{shortversion}}%
{}%
{Of these 70 programs, 52 are typable by our static semantics. }%
The language we used for testing was Caml Light, a pedagogical sublanguage of a precursor to OCaml, for which we have an existing translation to funcons from a previous case study~\cite{Churchill:15:ReusableComponents}.
We extended Caml Light with control operators, and specified the semantics of those operators as direct translations into the corresponding funcons presented in this paper.
The generated funcon programs were then tested by our prototype funcon interpreter, which directly interprets their \IMSOS{} specifications.
The suite of test programs, and our accompanying translator and interpreter, are available online~\cite{AdditionalMaterial}.

While the test programs demonstrated that we had successfully specified a control operator that behaves very similarly to the operator \op{control} described in the literature, they did not prove that we had specified exactly the same operator.
We addressed this in Section~\ref{sec:adequacy}, where we proved that our MSOS specification is equivalent to a conventional specification using a reduction semantics based on evaluation contexts (e.g. \cite{Felleisen:88:AbstractContinuations,Kameyama:08:TypedPromptControl}).

\ifthenelse{\boolean{shortversion}}%
{%
}
{
The ternary type used for lambda abstraction in our static semantics was inspired by our earlier treatment of dynamically scoped bindings in~\cite{Churchill:15:ReusableComponents}.
It remains as future work to unify these two approaches in \MSOS{}, with the aim of specifying a type of abstractions that is parameterised over \emph{any} contextual restrictions that should be discharged where an abstraction is applied, rather than where it is defined.
}

%% file: acknowledgement.tex
\paragraph{Acknowledgments:}
We thank Casper Bach Poulsen, Ferdinand Vesely and the anonymous
reviewers for feedback on earlier versions of this paper.  We also
thank Martin Churchill for his exploratory notes on adding evaluation
contexts to \MSOS{}, and Olivier Danvy for suggesting additional test
programs.  The reported work was supported by EPSRC grant
(EP/I032495/1) to Swansea University for the \textsc{PLanCompS}
project and by EU funding (Horizon 2020, grant 640954) to KU Leuven
for the \textsc{GRACeFUL} project.